\documentstyle[aps,prb,subfigure,epsfig,amsmath,amssymb]{revtex}
\begin{document}
\draft

\title{Analytical expressions for the spin-spin 
local-field factor and the spin-antisymmetric exchange-correlation kernel of
a two-dimensional electron gas}
\author{B. Davoudi$^{1,2}$, M. Polini$^1$, G. F. Giuliani$^3$ 
and M. P. Tosi$^1$}
\address{$^1$NEST-INFM and 
Classe di Scienze, Scuola Normale Superiore, I-56126 Pisa, Italy\\
$^2$Institute for Studies in Theoretical Physics and Mathematics, Tehran
19395-5531, Iran\\
$^3$ Physics Department, Purdue University, West Lafayette, Indiana
}
\maketitle
\vspace{0.2 cm}

\begin{abstract}
We present an analytical expression for the static many-body 
local field factor $G_{-}(q)$ of a homogeneous two-dimensional electron 
gas, which reproduces Diffusion Monte Carlo data and embodies the 
exact asymptotic behaviors at both small and large wave number $q$. 
This allows us to also provide a closed-form expression for the
spin-antisymmetric exchange 
and correlation kernel $K^{-}_{xc}(r)$ which 
represents a key input for spin-density functional studies of inhomogeneous
electronic systems.
\end{abstract}
\vspace{0.4 cm}

The static spin-spin response function 
$\chi_{\rm \scriptscriptstyle S}(q)$ of a paramagnetic electron gas (EG)
can be written in terms of the Lindhard function $\chi_0(q)$ by means of
the spin-antisymmetric many-body local field $G_-(q)$ through the relationship
\begin{equation}
\chi_{\rm \scriptscriptstyle S}(q)
=-\mu^{2}_{B}\,\frac{\chi_0(q)}{1+v_{q}G_-(q)\,\chi_0(q)} ~,
\end{equation}
where $\mu_{B}$ is the Bohr magneton.
Thus $G_-(q)$ is a fundamental quantity for the determination of many
properties of a general electron system. It is a key input, together
with the charge-charge local-field factor $G_+(q)$, in the
spin-density functional theory (SDFT) of the inhomogeneous electron
gas\cite{bp,gross} and in studies of 
quasiparticle properties (such as the effective mass and 
the effective Land\`e g-factor) in the electronic Fermi
liquid\cite{giuliani,pudalov}.   

For what concerns SDFT calculations, 
a common approximation to the unknown
exchange-correlation energy functional  
$E_{\rm \scriptscriptstyle xc}[n, \zeta]$ of electron density $n$ and
spin-polarization $\zeta$ appeals to its 
second functional derivatives, 
\begin{equation}
K^{\sigma \sigma'}_{\rm \scriptscriptstyle xc}({\bar n}, {\bar \zeta};
|{\bf r}-{\bf r}'|)\equiv 
\left.\frac{\delta^{2} E_{\rm
\scriptscriptstyle xc}[n, \zeta]}{\delta n_{\sigma}({\bf r}) \delta
n_{\sigma'}({\bf r}')}\right|_{n={\bar n},\, \zeta={\bar \zeta}}\, ,
\end{equation}
where ${\bar n}={\bar n}_{\uparrow}+{\bar n}_{\downarrow}$ 
is the average local density of the EG and 
${\bar \zeta}= ({\bar n}_{\uparrow}-{\bar n}_{\downarrow})/{\bar n}$ 
is the average local spin-polarization. The local
field factors $G_{\sigma \sigma'}(q)$ and  
the matrix of the exchange-correlation kernels 
are simply related in Fourier transform by
\begin{equation}\label{1}
{\widetilde K}^{\sigma \sigma'}_{\rm \scriptscriptstyle xc}(q)\equiv 
\int \!d^{d}r\,e^{-i\,{\bf
q}\cdot {\bf r}}\,K^{\sigma \sigma'}_{\rm \scriptscriptstyle xc}(r) 
=-v_{q}\,G_{\sigma \sigma'}(q)\, ,
\end{equation}
where $d$ is the dimensionality of the system and 
$v_{q}$ is the Fourier transform of the Coulomb potential
$e^{2}/r$. In the paramagnetic state the following relation holds
between the charge-charge 
and spin-spin local-field factors $G_{\pm}(q)$ and 
$G_{\sigma \sigma'}(q)$: 
\begin{equation}\label{rel}
G_{\pm}(q)=\frac{G_{\uparrow \uparrow}(q)\pm G_{\uparrow
\downarrow}(q)}{2}~.
\end{equation}
Thus, from Eq.~(\ref{1}) and (\ref{rel}), one obtains that the
spin-symmetric and the spin-antisymmetric exchange and correlation
kernels $K^{\pm}_{\rm \scriptscriptstyle xc}(r)$ are related to 
$G_{\pm}(q)$ by
\begin{equation}\label{5}
{\widetilde K}^{\pm}_{\rm \scriptscriptstyle xc}(q)\equiv
 \frac{{\widetilde K}^{\uparrow \uparrow}_{\rm \scriptscriptstyle
 xc}(q) \pm
 {\widetilde K}^{\uparrow \downarrow}_{\rm \scriptscriptstyle xc}(q)}{2}\equiv 
\int \!d^{d}r\,e^{-i\,{\bf
q}\cdot {\bf r}}\,K^{\pm}_{\rm \scriptscriptstyle xc}(r) 
=-v_{q}\,G_{\pm}(q)~.
\end{equation}
In what follows we shall only consider 
$K^{-}_{\rm \scriptscriptstyle xc}(r)$ in the case of
two spatial dimensions, with $d=2$ and $v_{q}=2 \pi e^{2}/q$. 
An analytical expression for $K^{+}_{\rm \scriptscriptstyle xc}(r)$
has already been given in Ref.~[$\!\!$ \onlinecite{bahman}]. 

A number of exact asymptotic properties of $G_{-}(q)$ 
in two dimensions are readily proven. In
particular\cite{SanGiu},
\begin{equation}\label{smallqminus}
\lim_{q\rightarrow 0}
G_{-}(q)= A_- \frac{q}{k_{F}}
\end{equation}  
with 
\begin{equation} 
A_-=\frac{1}{r_{s}\sqrt{2}}\,\left(1-\frac{\chi_{0}}{\chi}\right) ~,
\end{equation}
where $k_{F}=\sqrt{2 \pi n}=\sqrt{2}/r_{s} a_{B}$ is the Fermi wave
number, $r_{s}=\sqrt{\pi n a^{2}_B}$ is the usual EG density parameter
with $a_{B}$ the Bohr radius, $\chi_{0}=\mu^2_B\,m/\pi \hbar^{2}$ 
is the (Pauli) spin susceptibility of the ideal Fermi gas 
while $\chi$ is the spin susceptibility of the interacting system.
By making use of the thermodynamic definition of $\chi$ we can write
\begin{equation}\label{8}
\frac{\chi_{0}}{\chi}=1-\frac{\sqrt{2}}{\pi}\,r_s
+\frac{r^{2}_s}{2}\,\left.\frac{
\partial^{2} \epsilon_{c}}{\partial \zeta^{2}}\right|_{\zeta=0} 
\end{equation}
where $\epsilon_{c}$ is the correlation energy per particle in the
spin-polarized fluid. Unfortunately the correlation energy at finite spin
polarization is poorly known. By extrapolating 
the Diffusion Monte Carlo (DMC) 
data\cite{MCS} for $G_-(q)$  to $q=0$ and by observing from
Eq.~(\ref{8}) that
$A_-\rightarrow 1/\pi$ for $r_s \rightarrow 0$,  we
obtain the following parametrization of $A_-$ in the range $0\leq
r_s\leq 10$:
\begin{equation}
A_-=\frac{1}{\pi +1.4954\,r_s+0.3193\,\sqrt{r_s}}\, .
\end{equation}

The asymptotic behaviour of $G_{-}(q)$ at large $q$ is also known 
exactly\cite{SanGiu,holas},  
\begin{equation}\label{largeqminus}
G_{-}(q)\sim C_-\, \frac{q}{k_{F}}+B_-
\end{equation}
for $q \rightarrow \infty$,  where $C_-$ 
is proportional to the difference in kinetic energy between the
interacting and the ideal gas,
\begin{equation}
C_-=\frac{t-t_{0}}{2 \pi n e^{2}}\,k_{F}= 
-\frac{r_{s}}{2\sqrt{2}}\,\frac{d}{d
r_{s}}\Big(r_{s}\epsilon_{c}(r_{s})\Big)~.
\end{equation}
The correlation energy per particle $\epsilon_{c}(r_{s})$ 
in the paramagnetic fluid is available from Monte Carlo
data\cite{rap}. Finally  
$B_-=g(0)$, $g(0)$ being the value of the pair-correlation
function at the origin\cite{Marco}.

\begin{figure}[h!]
\centerline{\mbox{\psfig{figure=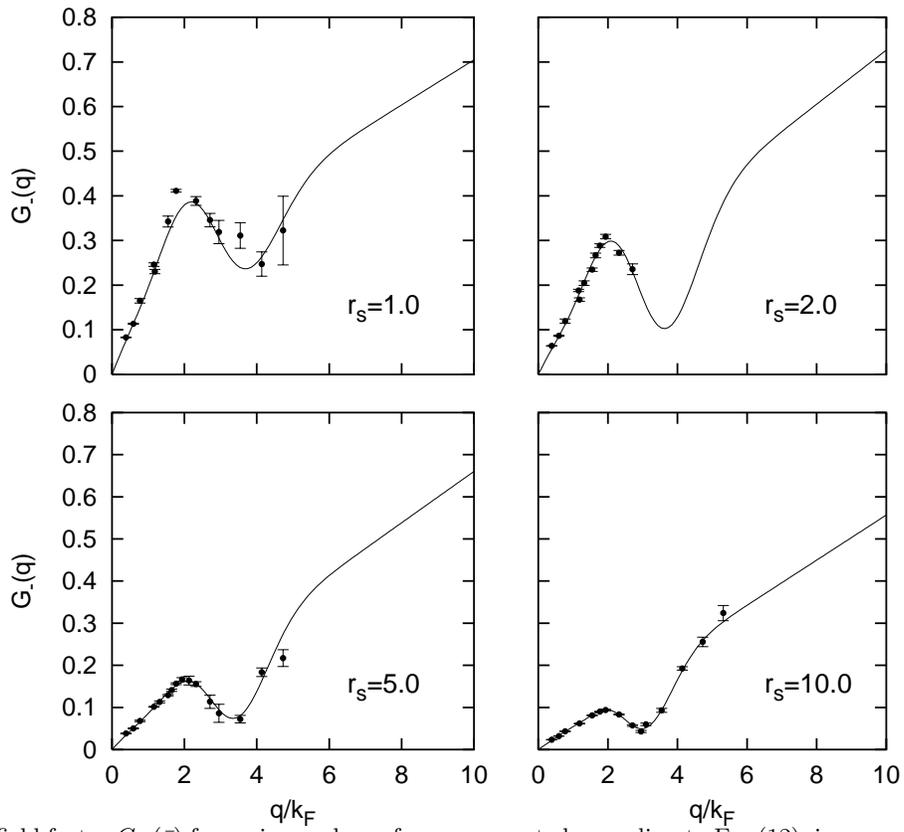, angle =0, width = 12 cm}}}
\caption{The local field factor $G_{-}({\bar q})$ for various values
of $r_{s}$ as computed according to Eq.~(\ref{gm}), in comparison
with the DMC data of Ref. 7}
\label{fig1}
\end{figure}

\begin{figure}[h!]
\centerline{\mbox{\psfig{figure=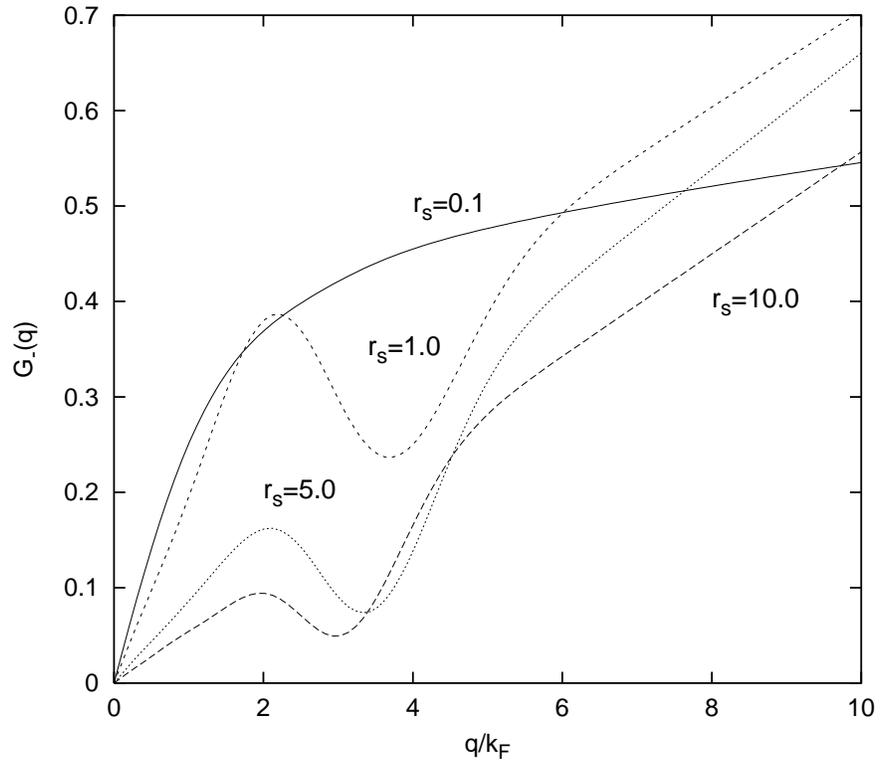, angle =0, width = 12 cm}}}
\caption{The local field factor $G_{-}(q)$ as from Eq.~(\ref{gm})
for various values of $r_{s}$.}
\label{fig2}
\end{figure}

In this work we fit the values of $G_{-}(q)$ originally obtained by
DMC in Ref.~[$\!\!$\onlinecite{MCS}]
in such a way as to obtain analytical expressions for both 
${\widetilde K}^-_{\rm \scriptscriptstyle xc}(q)$ and 
$K^-_{\rm \scriptscriptstyle xc}(r)$. 
The functional form previously employed\cite{bahman} for the
charge-charge local-field factor $G_+(q)$  
is also suitable for $G_{-}(q)$:
\begin{equation}\label{gm}
G_{-}({\bar q})=A_-\,{\bar q}\,\,
\left[\frac{\kappa_-(r_s)}{\sqrt{1+\left(A_-\,\kappa_-(r_s)\,{\bar
q}/B_-\right)^{2}}}+\left(1-\kappa_{-}(r_s)\right)\,
e^{-{\bar q}^{2}/4}\right] +
C_-\,{\bar q}\left(1-e^{-\beta_-{\bar q}^{2}}\right)+{\rm P}_-({\bar q})\,
e^{-\alpha_-\,{\bar q}^{2}}\, , 
\end{equation}
where ${\bar q}=q/k_F$, $\kappa_-(r_s)=\sqrt{1+0.0082\,r^2_s}$ and
${\rm P}_-({\bar q})$ is the polynomial
${\rm P}_-({\bar q})=h_2\,{\bar q}^{2}+h_4\,{\bar q}^{4}+h_6\,{\bar 
q}^{6}+ h_8 \, {\bar q}^{8}$.
The free parameters $\alpha_-$, $\beta_-$ and $h_{2},...,h_{8}$ 
are fitted so as to minimize the differences from the DMC numerical results. 
As for $G_+(q)$, it proves useful to have a continuous 
parametrization  of these parameters 
at least in the range $0\leq r_{s} \leq 10$. 
We propose the following:
\begin{eqnarray}\label{fittagm}
&&\alpha_-(r_{s})=
\exp{\left(-\frac{0.2231+81.2115\,(r_s/10)}{1+54.6665\,(r_s/10)+50.7534\,(r_s/10)^2}\right)}~,
\nonumber \\
&&\beta_-(r_{s})=0.8089-0.4025\,(r_s/10)^3-0.0941\,(r_s/10)^{1/2}~,
\nonumber \\
&& h_2 (r_{s})=
\exp{\left(-\frac{12.6262+20.9673\,(r_s/10)}{1+12.4002\,(r_s/10)}\right)}~,
\nonumber\\
&& h_4(r_{s})=
0.0531\,\Big(1-e^{-2.154\,r^2_s}\Big)-0.4984\,(r_s/10)^{3/2}
+0.4021\,(r_s/10)^2~, 
\nonumber \\
&& h_6 (r_{s})=-0.0076\,\Big(1-e^{-r_s}\Big)+0.0977\,(r_s/10)^{3/2}
-0.0726\,(r_s/10)^2~,\nonumber \\
&& h_8 (r_{s})=-0.0027\,(r_s/10)\, . 
\end{eqnarray}
In Figure~\ref{fig1} we compare the fit given by 
Eq.~(\ref{gm}) with the DMC data for $r_{s}=1,2,5$ and $10$. 
In Figure~\ref{fig2} 
we show the local field factor $G_{-}(q)$ as from
Eq.~(\ref{gm}) for various values of $r_{s}$.

We turn next to the evaluation of 
$K^{-}_{\rm \scriptscriptstyle xc}(r)$.
From Eq.~(\ref{5}) and (\ref{gm}),  
the expression of the spin-antisymmetric exchange-correlation
kernel in real space
(in Ryd) is readily obtained as
\begin{equation}\label{kxc}
K^{-}_{\rm \scriptscriptstyle xc}(r) = 
N_1 \frac{\delta^{(2)}({\bf r})}{k_F^2} 
+ N_2\,\frac{\exp{\Big(-B_-\,k_Fr/(A_-\kappa_-(r_s))\Big)}}{k_Fr}+
N_3\,e^{-(k_Fr)^2}+N_4\,e^{-(k_Fr)^2/4\beta_-}+
\sum_{n=1}^{4} N_{5,2n} F_{2n} (\alpha_-, k_Fr) ~,
\end{equation}
where $N_{1}=-4\,\pi\, \sqrt{2}\, C_-/r_{s}$, 
$N_{2}=-2\sqrt{2}\,B_-/r_{s}$, 
$N_3=-4\,\sqrt{2}\,A_-\,(1-\kappa_{-}(r_s))/r_{s}$,
$N_4=\sqrt{2}\,C_-/r_{s}\beta_-$ 
and $N_{5,n} = -2^\frac{3}{2}\,h_{n}\,/r_s$.
The function $F_n(\alpha, x)$ is given by
\begin{equation}
F_n (\alpha, x)=\int_{0}^{\infty}\,dy\,y^{n}\,{\rm
J}_{0}(xy)\,e^{-\alpha\,y^2}=\frac{1}{2}
\,\alpha^{-(1+n)/2}\,\Gamma(\frac{1+n}{2})\,\,_{1}{\rm
F}_{1}\left(\frac{1+n}{2};1; -\frac{x^2}{4\alpha}\right) 
\end{equation}
where $\Gamma(z)$ is Euler's Gamma function and 
$_{1}{\rm F}_{1}(a;b;z)$ is Kummer's function. 
In practice the function $F_{n}(\alpha, x)$ 
can be obtained via the recursive relation
\begin{equation}
F_{n+2}(\alpha, x)=-\frac{d \,F_{n}(\alpha, x)}{d\,
\alpha}~,\,\,\,\,F_{2}(\alpha,x)=\frac{\sqrt{\pi}}{16\,
\alpha^{5/2}}\,\left[(4\alpha-x^2)\,{\rm
I}_{0}\left(\frac{x^{2}}{8\alpha}\right)+x^2\, {\rm
I}_{1}\left(\frac{x^{2}}{8\alpha}\right)\right]\,e^{-x^{2}/8\alpha}
\end{equation}
where ${\rm I}_{n}(z)$ is the modified Bessel function of order
$n$. It is also useful to recall that $d {\rm I}_{0}(z)/dz={\rm
I}_{1}(z)$ and that $d {\rm I}_{1}(z)/dz={\rm I}_{0}(z)-{\rm
I}_{1}(z)/z$.

\begin{figure}[h!]
\centerline{\mbox{\psfig{figure=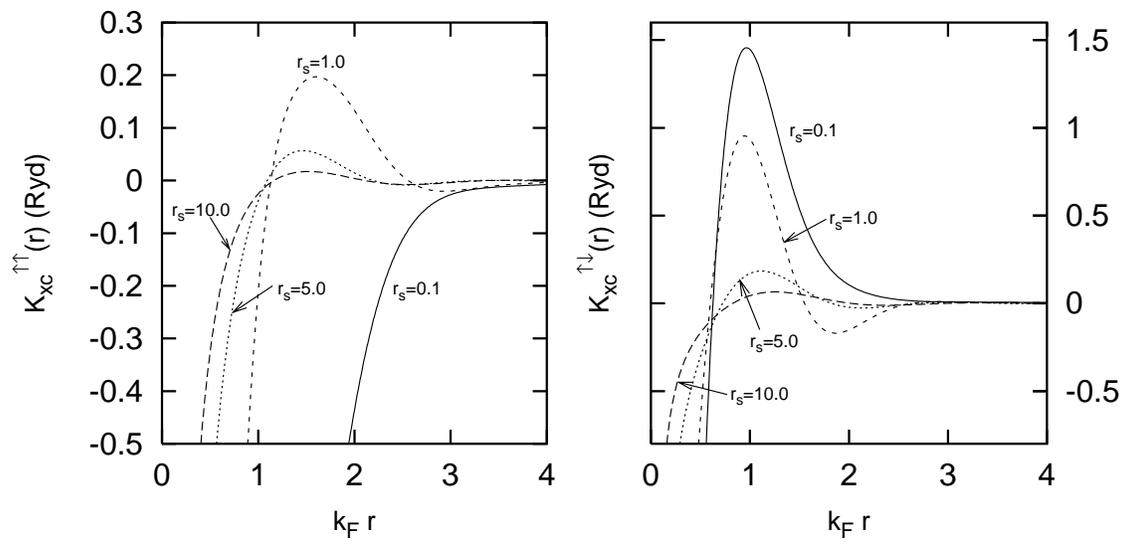, angle =0, width = 15 cm}}}
\caption{The matrix elements 
$K^{\uparrow \uparrow}_{\rm \scriptscriptstyle xc}(r)$ and 
$K^{\uparrow \downarrow}_{\rm \scriptscriptstyle xc}(r)$ of the
exchange-correlation kernel for various values of $r_{s}$.}
\label{fig3}
\end{figure}

In Figure~\ref{fig3} we show the matrix elements 
$K^{\sigma \sigma'}_{\rm \scriptscriptstyle xc}(r)$ of the 
exchange-correlation kernel for various values of $r_{s}$.
These kernels describe the local structure (Pauli-Coulomb hole) of the
paramagnetic electron fluid around an electron of a given spin. 
$K^{\uparrow \uparrow}_{\rm \scriptscriptstyle xc}(r)$, being the sum
of $K^{+}_{\rm \scriptscriptstyle xc}(r)$ and
$K^{-}_{\rm \scriptscriptstyle xc}(r)$, has the same behavior of these
two components, namely no structure at both small and high $r_s$. The
situation is different for 
$K^{\uparrow \downarrow}_{\rm \scriptscriptstyle xc}(r)$, which is
instead given by the difference of $K^{+}_{\rm \scriptscriptstyle xc}(r)$ and
$K^{-}_{\rm \scriptscriptstyle xc}(r)$. Although these two components
are structureless at small $r_s$, the difference 
between them around $k_F r \simeq
1$ is large and this is reflected in the large structure
shown in Figure~\ref{fig3}.

In conclusion, we have presented an analytic parametrization of the
local field factor entering the spin response of the two-dimensional
electron gas in the paramagnetic state, incorporating the known
asymptotic behaviors and giving an accurate representation of the
available quantum Monte Carlo data. We have obtained from it and from
our earlier results on the dielectric response\cite{bahman} analytic
expressions of the exchange-correlation kernels for
spin-density-functional calculations on inhomogeneous two-dimensional
electronic systems.

\acknowledgements
This work was partially supported by MURST through the PRIN 1999 program.
We are grateful to Dr. S. Moroni for fruitful 
discussions and for providing us 
with the results of the Diffusion Monte Carlo study. Three of us
(B. D., M. P. and M. P. T.) wish to thank the Condensed Matter Group
of the Abdus Salam International Center for Theoretical Physics in
Trieste for their hospitality during the final stages of this work.
%%%%%%%%%%%%%%%%%%%%%%%%%%%%%%%%%%%%%%%%%%%%%%%%%%%%%%%%%%%%%%%%%%%%%%%
%%%%%%%%%%%%%%%%%%%%%%%%%%%%%%%%%%%%%%%%%%%%%%%%%%%%%%%%%%%%%%%%%%%%%%%%%%%%%%

\end{document}